# PDBImages: A Command Line Tool for Automated Macromolecular Structure Visualization

Adam Midlik[1], Sreenath Nair[1], Stephen Anyango[1], Mandar Deshpande[1], David Sehnal[2,3], Mihaly Varadi[1,*], and Sameer Velankar[1,*]

[1] Protein Data Bank in Europe, European Molecular Biology Laboratory, European Bioinformatics Institute (EMBL-EBI), Wellcome Genome Campus, Hinxton, Cambridgeshire CB10 1SD, UK

[2] Biological Data Management and Analysis Core Facility, Centre for Structural Biology, CEITEC – Central European Institute of Technology, Masaryk University, Brno 62500, Czech Republic

[3] National Centre for Biomolecular Research, Faculty of Science, Masaryk University, Brno 62500, Czech Republic

* To whom correspondence should be addressed.

## Abstract

**Summary:** PDBImages is an innovative, open-source Node.js package that harnesses the power of the popular macromolecule structure visualization software Mol*. Designed for use by the scientific community, PDBImages provides a means to generate high-quality images for PDB and AlphaFold DB models. Its unique ability to render and save images directly to files in a browserless mode sets it apart, offering users a streamlined, automated process for macromolecular structure visualization. Here, we detail the implementation of PDBImages, enumerating its diverse image types and elaborating on its user-friendly setup. This powerful tool opens a new gateway for researchers to visualize, analyse, and share their work, fostering a deeper understanding of bioinformatics.

**Availability and Implementation:** PDBImages is available as an npm package from https://www.npmjs.com/package/pdb-images. The source code is available from https://github.com/PDBeurope/pdb-images.

**Contact:** mvaradi@ebi.ac.uk, sameer@ebi.ac.uk

## Introduction

Visualization of macromolecular structure data holds high importance for researchers seeking to explore and understand the intricate world of biomolecules (O'Donoghue *et al.*, 2010; Kozlíková *et al.*, 2017; Olson, 2018). While programmatic analyses of atomic coordinates can reveal valuable insights into biological processes and disease mechanisms, the power of viewing structures in three dimensions cannot be understated (Goddard and Ferrin, 2007; O'Donoghue, 2021). It is akin to providing the scientific community with a microscope to delve deeper into the atomic realm and observe molecular interactions in unprecedented detail.

A diverse range of tools currently exist for viewing macromolecular 3D structures, ranging from desktop applications such as PyMol (Schrödinger, LLC.) and ChimeraX (Pettersen *et al.*, 2021) to browser-based tools like JSMol (Hanson *et al.*, 2013) and Mol* (Sehnal *et al.*, 2021). Each of these tools offers unique advantages and perspectives. However, among these software suites, Mol* has carved out a niche as a powerful and popular tool for displaying macromolecule structure data, and has been adopted by the Protein Data Bank in Europe (PDBe) (Armstrong *et al.*, 2020), the AlphaFold Protein Structure Database (Varadi *et al.*, 2021), UniProt (The UniProt Consortium, 2023), InterPro (Paysan-Lafosse *et al.*, 2023), Ensembl (Cunningham *et al.*, 2022), and several other major data providers. Mol* also stands out for its ability to save a given view as a state file that can be reloaded to reproduce the same view. However, to create macromolecular images and/or state files the user must manually create the desired views (e.g. load the structure, apply colouring, zoom to their region of interest) and export them one by one. For creating a large number of images, this approach becomes extremely inefficient.

To facilitate and automate image generation, we have developed PDBImages, a performant and reusable open-source software tool that leverages the capabilities of Mol* in conjunction with the PDBe API (Armstrong *et al.*, 2020). PDBImages aims at fully automated generation of high-quality images for PDB entries and AlphaFold DB models, with an easy-to-use command-line interface and scalability. We utilize this tool internally to generate images displayed on the PDBe pages, but it can also be employed directly by the scientific community, facilitating the generation of molecular structure images for communication of scientific outcomes.

This application note provides an overview of PDBImages, detailing its functionality and instructions for usage, enabling users to generate high-quality molecular structure images efficiently.

## Implementation

PDBImages is a Node.js command-line application, written in TypeScript and building on the popular visualization library Mol* (Sehnal *et al.*, 2021). Its core functionality revolves around the ability to read atomic XYZ coordinates, construct predefined views of the macromolecular structures, and save these views as PNG images, Mol* state files, and caption files. To this end it employs Mol* in the browserless mode and *gl* rendering library. The implemented functionality is accessible via a straightforward, user-friendly command-line interface (Figure 1).

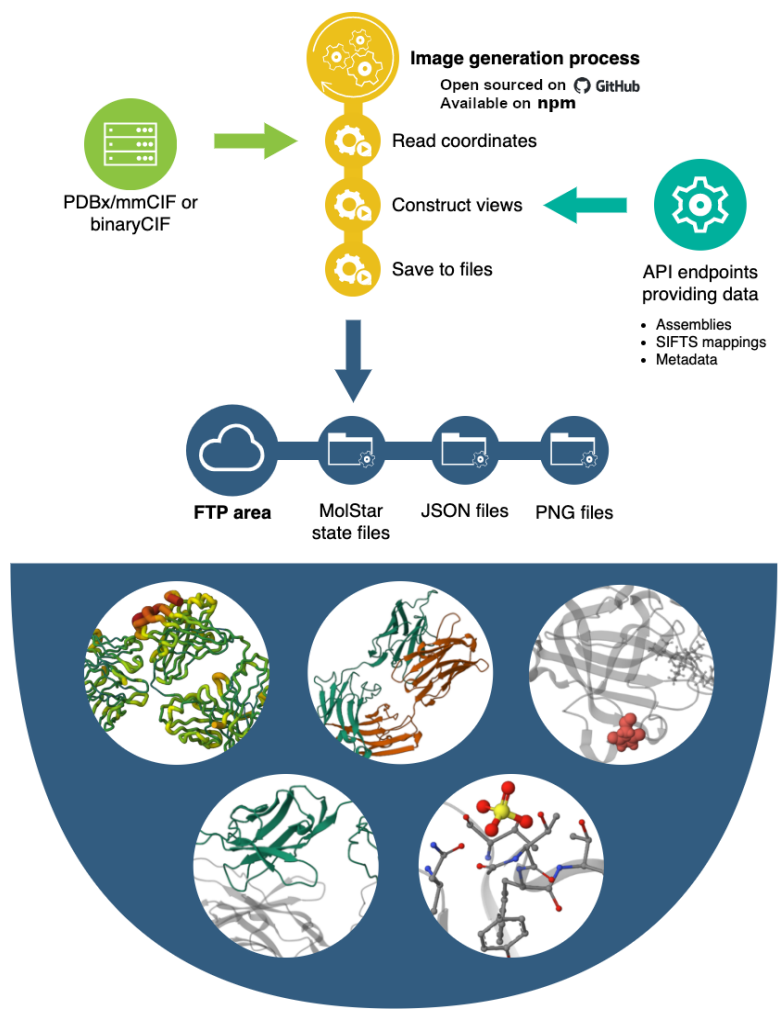

**Fig. 1.** Overview of the PDBImages tool.

PDBImages can work in two modes: the default *pdb* mode is suitable for PDB entries and custom structures, while the *alphafold* mode is dedicated to AlphaFold predicted models.

First, PDBImages reads the coordinate file of the processed structure, which may be a PDB entry, AlphaFold model, or a custom structure. The input structure file can be in the PDBx/mmCIF (.cif) or binary CIF (.bcif) format and can also be compressed with GZIP (.cif.gz, .bcif.gz). When the --*input* option is not specified, the input file will be automatically retrieved from PDBe (Armstrong *et al.*, 2020) or AlphaFold DB (Varadi *et al.*, 2021), depending on the selected mode.

PDBImages then renders individual images and saves them in the output directory. It provides nine distinct image types, each focused on a different aspect of the structure. Eight types apply in the *pdb* mode and one in the *alphafold* mode (see Table 1).

**Table 1.** Image types generated by PDBImages

| Mode | Image type | Description |
|---|---|---|
| pdb | entry | Show the complete deposited structure, coloured by chains and entities (chemically distinct molecules). For ensembles, show all models. |
| pdb | assembly | For each assembly listed in the mmCIF file, show the entire assembly, coloured by chains and entities. |
| pdb | entity | For each entity, show the preferred assembly with this entity highlighted (excluding the water entity). |
| pdb | domain | Highlight domain mappings from CATH (Sillitoe *et al.*, 2021), SCOP (Andreeva *et al.*, 2020), Pfam (Mistry *et al.*, 2020) and Rfam (Kalvari *et al.*, 2021) using SIFTS mappings (Dana *et al.*, 2019). |
| pdb | ligand | For each distinct non-polymer entity (excluding water), show this entity and its surroundings. |
| pdb | modres | For each distinct type of modified residue in the structure, show the preferred assembly with all instances of this modified residue highlighted. |
| pdb | bfactor | Show the deposited structure in putty representation, colour-coded by B-factor data (only applies to X-ray structures). |
| pdb | validation | Show the deposited structure colour-coded by structure quality data. |
| alphafold | plddt | Show the predicted structure colour-coded by the pLDDT confidence measure (Jumper *et al.*, 2021) (only applies to computationally predicted structures). |

Image types can be selected using the --*type* option, or by default, all applicable image types will be rendered. For each image, at least three files are saved: the rendered image itself (in PNG format, possibly in multiple resolutions), a JSON file with the image caption, and a Mol* state file (MOLJ format, which can be loaded into Mol* to reproduce the same view, for example by dragging the state file and dropping it in a browser window with Mol* Viewer).

Some image types require additional input data; these are automatically fetched from the PDBe API (Armstrong *et al.*, 2020) (see Table 2). As this feature is only relevant for PDB entries, for other structures (not deposited to PDB) it can be disabled using the --*no-api* option.

Orientation of the visualized structure plays a crucial role, as an improperly selected orientation can lead to the occlusion problem – the parts of the structure closer to the viewer will hinder the visualisation of the structure farther away (Heinrich *et al.*, 2014). While this cannot be completely avoided, we minimize the chance of occlusion by calculating the principal component analysis (PCA) of the atomic coordinates and aligning the PCA axes to the screen axes, or "laying the structure flat against the screen". Additional rules ensure that the original orientation of the structure does not affect (flip) the resulting orientation. This new handy feature has already been integrated into the Mol* Viewer itself (*Orient Axes* option under the *Reset* button).

**Table 2.** Optional data retrieved from the PDBe API

| Retrieved information | PDBe API endpoints |
|---|---|
| Entity names | /pdb/entry/molecules/{id} |
| Preferred assembly information | /pdb/entry/summary/{id} |
| Modified residues | /pdb/entry/modified_AA_or_NA/{id} |
| Domain mappings from SIFTS | /mappings/{id}, /nucleic_mappings/{id} |
| Validation data | /validation/residuewise_outlier_summary/entry/{id} |

{id} is to be replaced by the PDB identifier.

This optimal orientation is referred to as the front view. Some image types are additionally rendered in the side view and top view, with arrows in the left bottom corner indicating the PCA axes (this can be adjusted by the *--view* and *--no-axes* options).

As the last step, PDBImages creates two summary files: the first, *{id}_filelist*, is a simple list of created images; the second, *{id}.json*, also provides image captions and other metadata and has the images structured into sections by image type.

Much more detailed instructions can be found in the PDBImage documentation (https://github.com/PDBeurope/pdb-images#pdbimages).

## Availability

PDBImages is released as an npm package (https://www.npmjs.com/package/pdb-images) and can be installed using the npm package manager (requires Node.js 18 or higher). In this way, it can be used as a standalone command-line application but can also be easily incorporated into more complex workflows. The source code for PDBImages is publicly available under Apache 2 licence from the PDBe GitHub repository at https://github.com/PDBeurope/pdb-images. We encourage contributions from the scientific community to further improve and expand the capabilities of PDBImages.

PDBImages will run seamlessly and utilize GPU for rendering on Linux, Mac, and Windows personal computers. For running in Linux environments without a running X-server (like large computing infrastructures), we provide a Docker image using X emulator Xvfb; however, this will not utilize GPU (https://hub.docker.com/r/pdbegroup/pdb-images).

PDBImages has been used to generate images displayed on the PDBe pages (https://www.ebi.ac.uk/pdbe/) since August 2023. For any PDB entry, all generated images can be downloaded in various resolutions (1600x1600, 800x800, 200x200, and 100x100 pixels, plus the Mol* state file). The list of available images for an entry can be obtained from the summary files (*https://www.ebi.ac.uk/pdbe/static/entry/{id}_filename* or *https://www.ebi.ac.uk/pdbe/static/entry/{id}.json*, where *{id}* stands for the PDB ID of interest).

The full URL of a specific file can then be obtained by combining the base URL, filename (retrieved from either of the summary files), and file suffix (retrieved from the JSON summary file), as demonstrated by this example:

https://www.ebi.ac.uk/pdbe/static/entry/
+ 1tqn_bfactor
+ _image-800x800.png
= https://www.ebi.ac.uk/pdbe/static/entry/1tqn_bfactor_image-800x800.png

Images for new and updated PDB entries are published simultaneously with the weekly PDB release.

## Acknowledgements

We would like to thank Alexander S. Rose, Sebastian Bittrich, and Jesse Liang, who contributed the code in the core Mol* library allowing it to run in the browserless mode.

## Funding

This work was supported by Biotechnology and Biological Sciences Research Council/National Science Foundation funding [BB/W017970/1, PI: S. Velankar; DBI-2129634, PI: S. K. Burley]; Wellcome Trust [218303/Z/19/Z to S. Velankar]; European Molecular Biology Laboratory – European Bioinformatics Institute; Czech Science Foundation [22-30571M to D. Sehnal]; and Ministry of Education, Youth and Sports of the Czech Republic [LM2023055 to D. Sehnal].

*Conflict of Interest:* none declared.